\documentclass{aac}

\usepackage[numbers]{natbib}
\usepackage{rotating}
\usepackage{graphicx}

\begin{document}

\title{Two-Pulse Direct Laser Acceleration in a Laser-Driven Plasma Accelerator}

\author[aff1]{Brant Bowers}
\author[aff1]{Max LaBerge}
\author[aff2]{Alexander Koehler}
\author[aff2]{Jurjen Couperus Cabada\u{g}}
\author[aff2]{Yen-Yu Chang}
\author[aff2]{Patrick Ufer}
\author[aff2]{Michal \v{S}m{\'i}d}
\author[aff2]{Arie Irman}
\author[aff3]{Tianhong Wang}
\author[aff1]{Aaron Bernstein}
\author[aff3]{Gennady Shvets}
\author[aff2,aff4]{Ulrich Schramm}
\author[aff1]{Michael Downer\corref{cor1}}

\affil[aff1]{Physics Department, The University of Texas-Austin, Austin, Texas 78712, USA}
\affil[aff2]{Institute of Radiation Physics, Helmholtz-Zentrum Dresden-Rossendorf, 01328 Dresden, Germany}
\affil[aff3]{School of Applied and Engineering Physics, Cornell University, Ithaca, New York 14850, USA}
\affil[aff4]{Technische Universit\"{a}t Dresden, 01062 Dresden, Germany}
\corresp[cor1]{downer@physics.utexas.edu}

\maketitle

\begin{abstract}
We present methods and preliminary observations of two pulse Direct Laser Acceleration in a Laser-Driven Plasma Accelerator. This acceleration mechanism uses a second co-propagating laser pulse to overlap and further accelerate electrons in a wakefield bubble, increasing energy at the cost of emittance when compared to traditional laser wakefield acceleration (LWFA). To this end, we introduce a method of femtosecond scale control of time delay between two co-propagating pulses. We show energy enhancement when the separation between the two pulses approaches the bubble radius. 
\end{abstract}

\section{INTRODUCTION}
Laser Wakefield Acceleration (LWFA) has progressed and diversified significantly since it was first proposed in $1979$ \cite{tajima1979laser}. This technique has traditionally relied on the ponderomotive force from tightly focused and compressed laser pulses to drive fast plasma waves which then transfer energy to the injected electrons. These waves can support accelerating fields $E_x[\hbox{V/cm}] \propto \sqrt{n_e [\hbox{cm}^{-3}]}$, where $n_e$ is the electron density, exceeding $1$ GV/cm \cite{faure2006controlled,esarey2009physics,downer2018diagnostics}, several orders of magnitude larger than modern radio-frequency accelerators \cite{behnke2013international}. The most common variation of this method is blowout or bubble regime LWFA where an ultrashort and high intensity, $a_0 > 1$, laser pulse in a sub-critical density plasma drives a near spherical electron-depleted region with a nearly linear radial electric field \cite{faure2004laser,geddes2004high,mangles2004monoenergetic}. A great deal of research has focused on manipulating the injection mechanism used to introduce these electrons into the acceleration process ranging from  the Self-Injection used in the first blowout LWFAs, to Self-Truncated Ionization Injection \cite{zeng2014self,couperus2017demonstration,irman2018improved}, to Shock Front Injection, \cite{buck2013shock} to Trojan Horse Injection \cite{deng2019generation}, to Downramp Injection \cite{schmid2010density}. However, work on controlling the evolving acceleration process has largely focused on implementing plasma channels \cite{tsung2004near,leemans2006gev}. We discuss the implementation of a secondary laser pulse to manipulate the electrons as they accelerate. In this proceeding, we demonstrate a method of femtosecond control of the timing between two co-propagating laser pulses.\\

The major benefits of implementing controllable direct laser acceleration (DLA) in an LWFA system are threefold. First, we can directly transfer energy from the DLA pulse to the accelerating electrons through a $\vec{v} \times \vec{B}$ force coupled to their betatron oscillation. Second, we can increase the dephasing length of the accelerating electrons by adding transverse momentum, relativistically reducing their longitudinal velocity. Thirdly, we can directly impact the wiggler strength experienced by the electrons in the bubble. Among other advantages, this allows us to enhance the x-ray energy and flux of the resulting betatron radiation without the need to increase the number or energy of the accelerating electrons. In this proceeding, we will focus only on the energy transfer as it is a good first indicator of successful overlap between the DLA pulse and the accelerating electrons. \\

Some past works have examined the impact of direct laser acceleration from the tail of a drive pulse on an accelerating electron bunch through both simulation \cite{shaw2016estimation} and experiment \cite{cipiccia2011gamma,shaw2018experimental}. In these studies, DLA was controlled by increasing the pulse length of the drive pulse, thereby increasing the amplitude of the pulse tail that interacted with the accelerating electron bunch. This experimental study is based on existing simulation and theory \cite{zhang2015synergistic,zhang2016laser,zhang2018effects,wang2019direct}. We seek to make the DLA process more controllable by using a second independent pulse to control electrons in a wake driven by the first pulse.  \\

The motivation of this work is to develop an independent technique to enhance LWFA electron energy beyond the traditional dephasing limit. The enhanced-energy bunch could then drive a Plasma Wakefield Accelerator (PWFA) \cite{litos2014high, martinez2019hybrid}, while the betatron x-ray output is useful for imaging applications \cite{kneip2011x,cole2015laser}.\\

\section{Experimental Methods}
To separate a DLA pulse of desired energy and controlled delay $\Delta T$ from the LWFA drive pulse, we inserted a segmented flat circular mirror into the  delivery line of the DRACO laser pulse at HZDR.  A central circular portion of this mirror (radius $2.5$ cm) reflected the center of the DRACO \cite{schramm2017first} pulse, directing it toward the gas jet to drive a LWFA.  The outer annulus of the segmented mirror reflected the remaining outer portion of the DRACO pulse, directing it also toward the gas jet to serve as a DLA pulse. \\
\begin{figure}[t]
    \centering
    \centerline{\includegraphics[width=400pt]{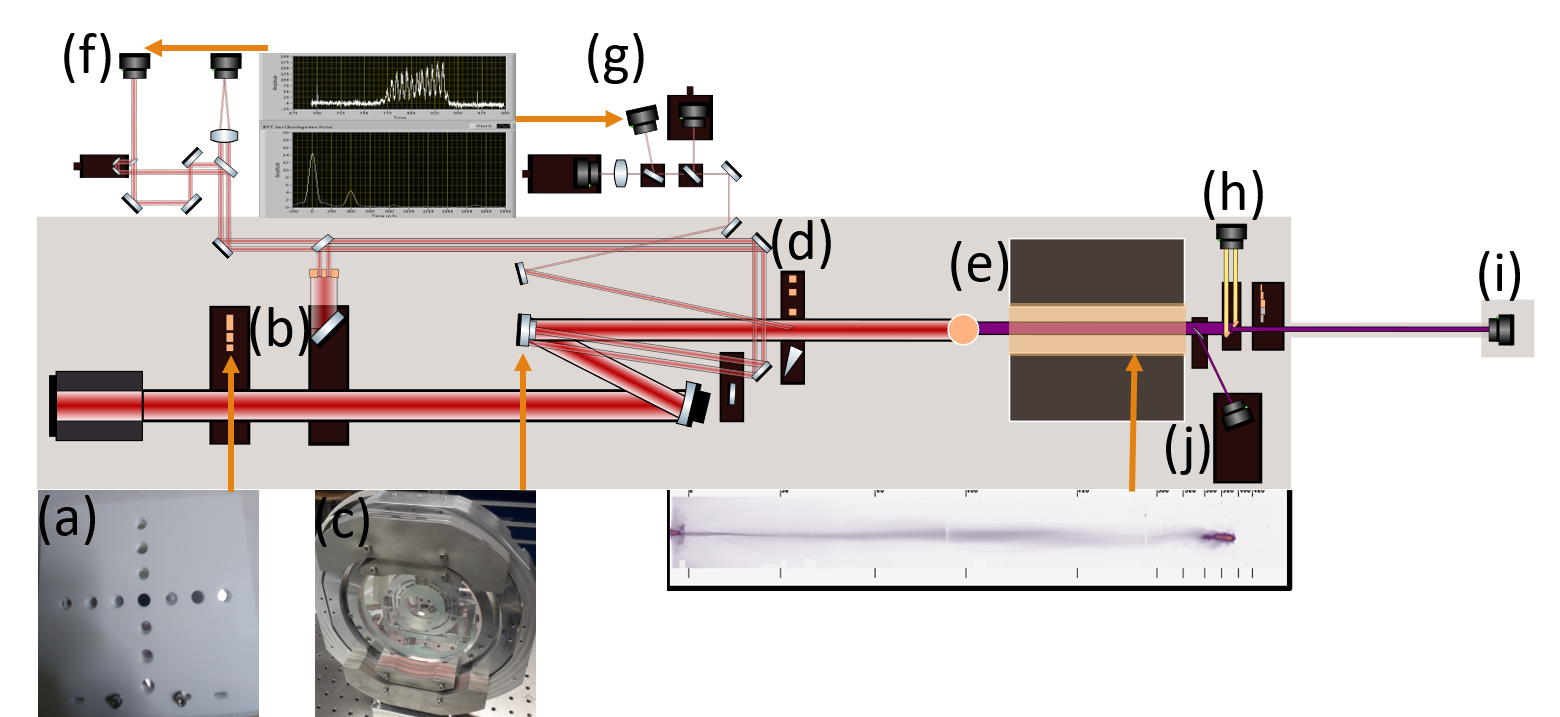}}
    \caption{A diagram of the target chamber setup. The DRACO laser enters from the left, then it can interact with (a) the dot mask, used to choose which portions of the beam to allow through for absolute delay diagnostics. A removable pickoff mirror (b) diverts the centermost portion of the beam for use in a relative time delay diagnostic. The beamline reflects from an OAP and then (c) our segmented mirror which adds a well-controlled time delay to the outer part of the pulse compared to the inner through sub-micron translations of its outer annulus. At (d) we can insert either ceramic masks that isolate the inner or outer pulse or a wedge that redirects the beamline for the absolute time diagnosis. The beam focuses on a 3 mm cylindrical gas jet \cite{couperus2016tomographic} (orange circle) where the interaction takes place. Afterwards, the electrons are diagnosed by (e) a dipole and four scintillating screens that resolve the electron charge, energy, and angular distribution in the laser polarization direction. After this, we diagnose x-rays with a x-ray scintillator (h),  pixel counting x-ray camera \cite{kohler2016single} (i), and HOPG spectrometer \cite{vsmid2017highly} (j). Our relative time delay calibrations are conducted at (f) through spectral interferometry and our absolute time delay calibrations are conducted at (g) with a separate spectral interferometer.}
\end{figure}
 \\
\begin{figure}[t]
    \centering
    \centerline{\includegraphics[width=400pt]{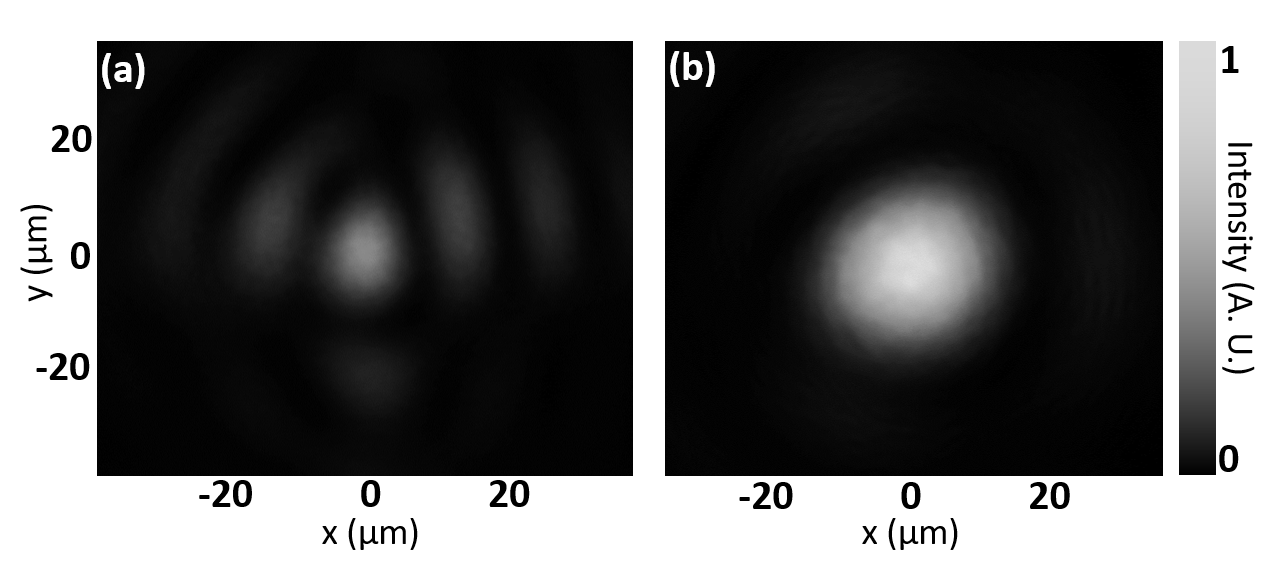}}
    \caption{Beam Profile of the (a) DLA pulse and (b) Drive pulse at the vacuum focus. Obtained through the use of the matched ceramic masks described in Fig. 1.}
\end{figure}
We utilized the experimental setup depicted in Fig. 1 in order to re-measure the added time delay $\delta t$ added or removed by our segmented mirror and to check both beam profiles between shot groups. The delay $\delta t$ tended to drift several femtoseconds in the negative direction over a group of $~20$ shots. This was likely due to feedback created from radiation generated on shot. This behaviour was eliminated by implementing a remote switch to disconnect the motor controller from the motor before taking shots. This source of error is however present in some data shown.

The beam profiles shown in Fig. 2 remained stable throughout the experiment. Through further calibrations, we determined the Drive pulse to have $E_1 = 2.5$ J and $\tau_{FWHM1} = 30$ fs and the DLA pulse to have $E_2 = 0.25$ J and $\tau_{FWHM2} = 40$ fs. Because the DLA pulse was selected from the outside of the fully compressed beam and compression was optimized for the pulse center, we see significantly more variability in the duration of the DLA pulse with $ 28\, \hbox{fs} < \tau_{FWHM2} < 50\, \hbox{fs}$ for more than $90\%$ of the measurements.

\begin{figure}[t]
    \centering
    \centerline{\includegraphics[width=400pt]{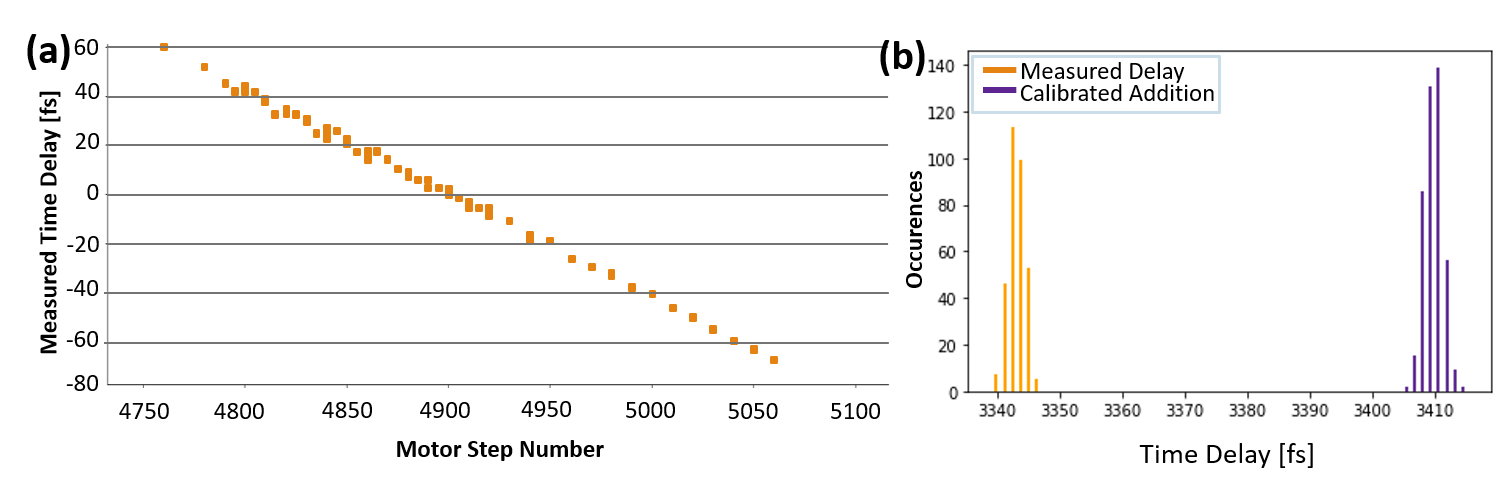}}
    \caption{(a) Measurements representing 3 scans over $40\, \mu$m demonstrating the repeatability of our closed-loop motor. We compare the measured time delay from our spectral interferometry setup to the step number readout on the Motor Controller software. (b) Measurements of the time delay added from our calibrated ND filters and the measured time delay at focus in vacuum with the dot mask in place. The difference represents 65fs delay between Drive and DLA pulses.}
\end{figure}
For relative time delay measurements, we used two methods. The first utilized the spectral interferometer shown in Fig. 1(f). Between shot groups the laser was switched from full power to partial power and higher rep rate, then the pointing of beams reflected from the inner and outer components of the segmented mirror was optimized. Then we verified the time delay offset of our spectral interferometer using a flat reference mirror. We would compare this to the measured time delay off our segmented mirror which used the same beam path with the reference mirror pulled to the out position. In the second method we used the spectral interferometry technique from the first method and scanned from $+20\,\mu$m to $-20\, \mu$m we repeated this in both directions. Figure 3(a) shows the results, which demonstrated that the correlation between nominal mirror movement and measured delay was both accurate and reproducible. This told us that we could rely on our closed loop motor to be accurate to within about $2.5$ fs of our intended time delay.\\

In order to establish our true time delay $\Delta T$ we measured the intrinsic time delay between the inside and outside of our compressed pulse $\Delta T_0$ in shot conditions. This was done by utilizing the dot mask in Fig. 1(a) with a pair of ND filters meant to even the intensity between the inner (Drive) and outer (DLA) parts of the beam while adding a known time delay between them. The beamlets left over after this mask then propagate through the system as they would during a shot until they get to the mask location in Fig. 1(d) where a wedge picks them off to travel to a spectral interferometer at Fig. 1(g).  There we measure the true current delay and compare it to the delay added by the ND filters. This comparison is shown in Fig. 3(b) and was measured to be close to $\Delta T_0 = 65$ fs meaning the DLA pulse inherently arrived $65$ fs after the drive pulse before adding or removing any time delay with our segmented mirror.

\subsection{Preliminary Results}

\begin{figure}[t]
    \centering
    \centerline{\includegraphics[width=400pt]{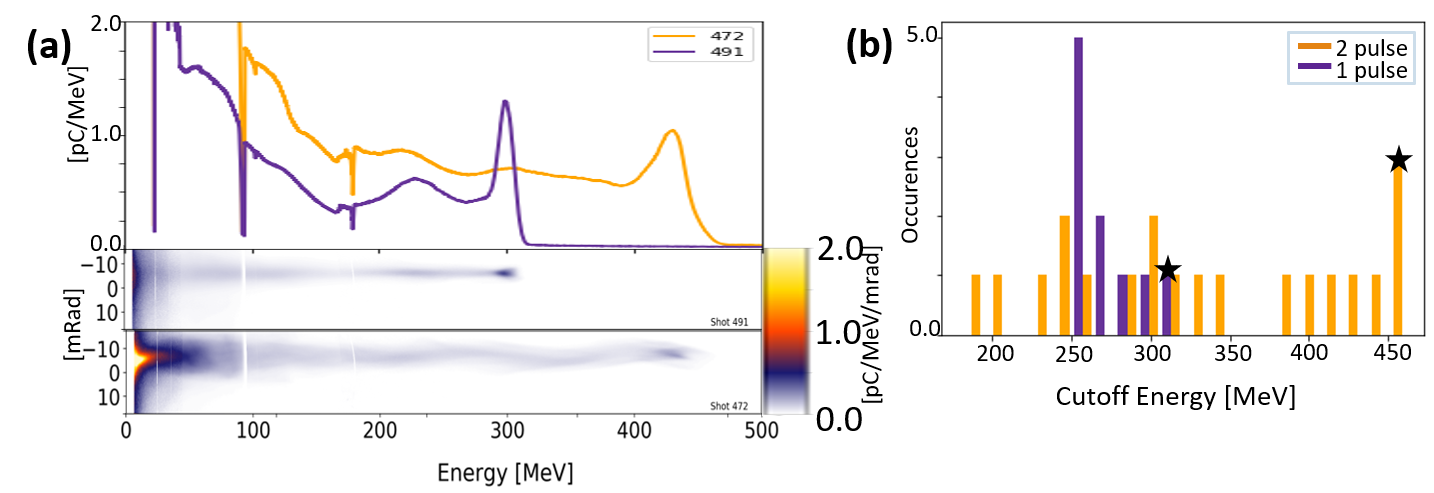}}
    \caption{(a) Raw data and energy lineout for the highest energy shots at both 2-pulse  $\Delta T = \delta t +  \Delta T_0 = 45$ fs  delay in orange and 1-pulse in purple. (b) Histogram of cutoff energies in both sets. The stars represent the data shown in (a).}
\end{figure}

We observed enhancement in the electron energy as a result of this DLA pulse. Fig. 4 shows results obtained in plasma of electron density $n_e = 4.0 \times 10^{18} \hbox{cm}^{-3}$, for which bubble radius is $~ 12\, \mu$m, the separation of two co-propagating pulses delayed by $\Delta T \sim  40$ fs.  This energy enhancement exceeded $100$ MeV in our best results [see Fig. 4(a)], but varied from shot to shot as shown in Fig. 4(b). We tentatively attribute these variations to compression instability in the DLA pulse and to the gradual wavelength scale motor motion that occurred as a result of unwanted feedback. This motion would change the way the tail of the drive pulse and the bulk of the DLA pulse interfere, changing the effective intensity profile that the accelerating electrons experience.  These variations notwithstanding, the energy enhancements were observed at $\Delta T$ corresponding to overlap of the DLA pulse with accelerating electrons, and exceeded normal shot-to-shot fluctuations of electron energy observed with only the drive laser pulse.

\section{ACKNOWLEDGMENTS}
This work was supported by U.S. DoE grant DE-SC0011617. M.D. acknowledges additional support from the Alexander von Humboldt Foundation.


\nocite{*}
\bibliographystyle{aac}%
\bibliography{aac2020_latex}%

\end{document}